# A pilot study examining transcranial photobiomodulation therapy intervention in college students with insomnia

**Jiangshan He[a,b,c], Lianghua Zhang[a], Dan Liang[a], Xiaoyu Wang[a], Tianyi Luo[a], Haoda Wang[a], Ziqi Ren[a], Mingzhe Jiang[d], Lei Zheng[e], Qiyuan Cheng[f,*], Hui Xie[a,b,c,*], Xueli Chen[a,b,c,d,g,*]**

[a] Center for Biomedical-photonics and Molecular Imaging (BMI), Advanced Diagnostic-Therapy Technology and Equipment Key Laboratory of Higher Education Institutions in Shaanxi Province, School of Life Science and Technology, Xidian University, Xi'an, Shaanxi 710126, China

[b] Institute of Interdisciplinary Artificial Intelligence (*i*AI), Faculty of Infor-X, Xidian University, Xi'an, Shaanxi 710126, China

[c] Xi'an Key Laboratory of Intelligent Sensing and Regulation of trans-Scale Life Information, School of Life Science and Technology, Xidian University, Xi'an, Shaanxi 710126, China

[d] Bi-optoelectronic-integration and Medical Instrumentation (BMI) Laboratory, Guangzhou Institute of Technology, Xidian University, Guangzhou, Guangdong 510555, China

[e] Macau University of Science and Technology, School of Business, Macau, China

[f] Medical Engineering Department, Shandong Provincial Hospital affiliated to Shandong First Medical University, Jinan 250021, China

[g] Chongqing Institute for Brain and Intelligence, Guangyang Bay Laboratory, Chongqing, 400064, China

* *Correspondence:* Xueli Chen, Email: xlchen@xidian.edu.cn; Hui Xie, E-mail: hxie@xidian.edu.cn; Qiyuan Cheng, Email: chengqy@sdu.edu.cn.

Dear Editor,

College students commonly report insufficient sleep and poor sleep quality, with ~30% meeting insomnia criteria, posing significant threats to their physical growth, cognitive development, and overall well-being, as well as imposing a substantial economic burden on society [1]. The hyperarousal model of insomnia [2] emphasizes that hyperarousal across cognitive, emotional, and physiological domains mutually reinforces one another. Neuroimaging studies have further identified prefrontal hypoactivity as a key neural substrate underlying these dysfunctional cognitions and elevated arousal, reflecting a failure of top-down modulatory control over both limbic reactivity [3] and brainstem arousal nuclei [4]. Moreover, transcranial photobiomodulation (tPBM) therapy targeting the prefrontal cortex has demonstrated therapeutic efficacy across neuropsychiatric disorders with insomnia comorbidities [5,6], providing preliminary support for its application in insomnia. However, the neuro mechanisms underlying tPBM's therapeutic effects on insomnia remain to be elucidated.

In this study, a randomized, single-blinded, assessor-blinded, sham-controlled experimental design was conducted. The study was approved by the institutional review board of the local institutes, and all participants voluntarily signed an informed consent before participation. 37 insomnia participants who met PSQI>5 criteria were enrolled and randomly assigned to tPBM ($n = 19$, 7 Female, $21.0 \pm 1.764$ years) or sham ($n = 18$, 7 Female, $20.0 \pm 0.686$ years) by random number.

Seven tPBM or sham consecutive sessions were delivered for 10 minutes per day. 10 Hz tPBM therapy was administered using a 980-nm laser in a 4-cm diameter circle (Model Aurora, Wuhan Jin Laser Medical Technology Co., Ltd., Wuhan, China) and located at the AF8 site in Figure 1(a). Multimodal assessment domains included behavioral scales (PSQI, ISI, sleep diary, PANAS), fearful face 2-back task, and EEG recording (eye-closed resting-state and ERP). Data were collected and compared between the baseline (week 0), post (week 1), follow-up 1 (week 2), and follow-up 2 (week 3) in Fig. 1(b). Mediation analysis elucidated the neurophysiological pathway underlying tPBM therapy efficacy, while the drift diffusion model was used to characterize how top-down modulation shapes the decision process under fearful face-induced emotional interference. Further experimental details are provided in the Supplementary material.

In Figure 1(b), PSQI of tPBM group showed sustained improvements with large effect sizes (post: p_bonf = 0.0038, d = 0.93; follow-up 1: p_bonf = 0.0027, d = 1.03; follow-up 2: p_bonf = 0.0002, d = 1.27). In Figure 1 (c), ISI of tPBM group showed reductions from baseline at all time points, with increasing efficacy (post: p_bonf = 0.0022, d = 0.700; follow-up 1: p_bonf = 0.0016, d = 0.723; follow-up 2: p_bonf = 0.0003, d = 1.112). In Figure 1 (d), relative delta power in the frontal region of tPBM group showed a sustained decrease (post: p_bonf = 0.005, d = 0.661). In Figure 1 (e), relative alpha power in the frontal region of tPBM group showed an increase (post: p_bonf = 0.011, d = 0.520). In Figure 1 (f) and (g), the mediation analysis revealed a significant indirect effect of the relative EEG power density compare with the baseline (f: indirect effect a × b = −0.940, p = 0.024; direct effect c' = −0.270, p = 0.785. g: indirect effect a × b = −0.830, p = 0.012; direct effect c' = 0.362, p = 0.713). The mediation analysis demonstrated that tPBM therapy influenced sleep disturbance through changes in relative EEG power density in the frontal region.

In Figure 1 (h), the tPBM group showed a trend of improvement in PANAS scores, but this change did not reach statistical significance. Figure 1 (i), drift rate $v$ in both groups improved, for the tPBM group (post: p_bonf = 0.002, d = 1.062; follow-up 1: p_bonf = 0.003, d = 0.992) and for the sham group (post: p_bonf = 0.004, d = 0.969; follow-up 1: p_bonf = 0.020, d = 0.730). Figure 1 (j), boundary separation $a$ of tPBM group showed a sustained decrease (post: p_bonf = 0.002, d = 1.111; follow-up 1: p_bonf = 0.013, d = 0.785). Figure 1 (k), a significant cluster-based reduction in non-target P300 amplitude was observed in the tPBM group compared to sham controls post-intervention, compare with the baseline (296.875 to 343.750 ms, p_cluster=0.015, Cohen's d=0.681), this neurophysiological improvement remained stable at the post (328.125 to 398.438 ms, p_cluster=0.005, Cohen's d=1.200) and follow-up 1 (195.312 to 242.188 ms, p_cluster=0.010, Cohen's d=0.817; 257.812 to 437.500 ms, p_cluster=0.001, Cohen's d=0.901) assessment. In the fearful face 2-back task, drift rate $v$ in both groups improved, which mainly reflects task-related practice. Beyond this shared practice effect, tPBM therapy enhanced neural efficiency optimization in the decision process by optimizing decision strategy (decreased boundary separation $a$) and reducing the attentional resource cost of non-target interference (decreased P300 amplitude).

This single-blind, sham-controlled study demonstrates that tPBM therapy at the right prefrontal cortex showed preliminary evidence of alleviating insomnia and improving neural efficiency in decision process with sustained benefits. Significant sleep disturbance improvements by tPBM therapy were evidenced by large effect size reductions in PSQI and ISI scores that persisted up to 2 weeks post-treatment. Notably, EEG changes preceded behavioral improvements, and the mediating pathway from EEG to sleep outcomes emerged at follow-up 1 rather than immediately post-treatment, suggesting that the neurophysiological mechanism underlying tPBM's efficacy unfolds with a temporal lag relative to its behavioral expression. tPBM therapy did not alter subjective mood or basic processing speed, but selectively optimized the quality of attentional control by refining decision thresholds and reducing the attentional cost of interference suppression [7]. These findings provide neuromechanistic evidence supporting tPBM as a promising intervention for insomnia. There are still limits. The follow-up period (up to 2 weeks post-treatment) is insufficient to conclude about durability. Future trials should include monthly follow-up assessments with objective sleep measures. Besides, tPBM therapy was well-tolerated in this sample, potentially worth further investigation.

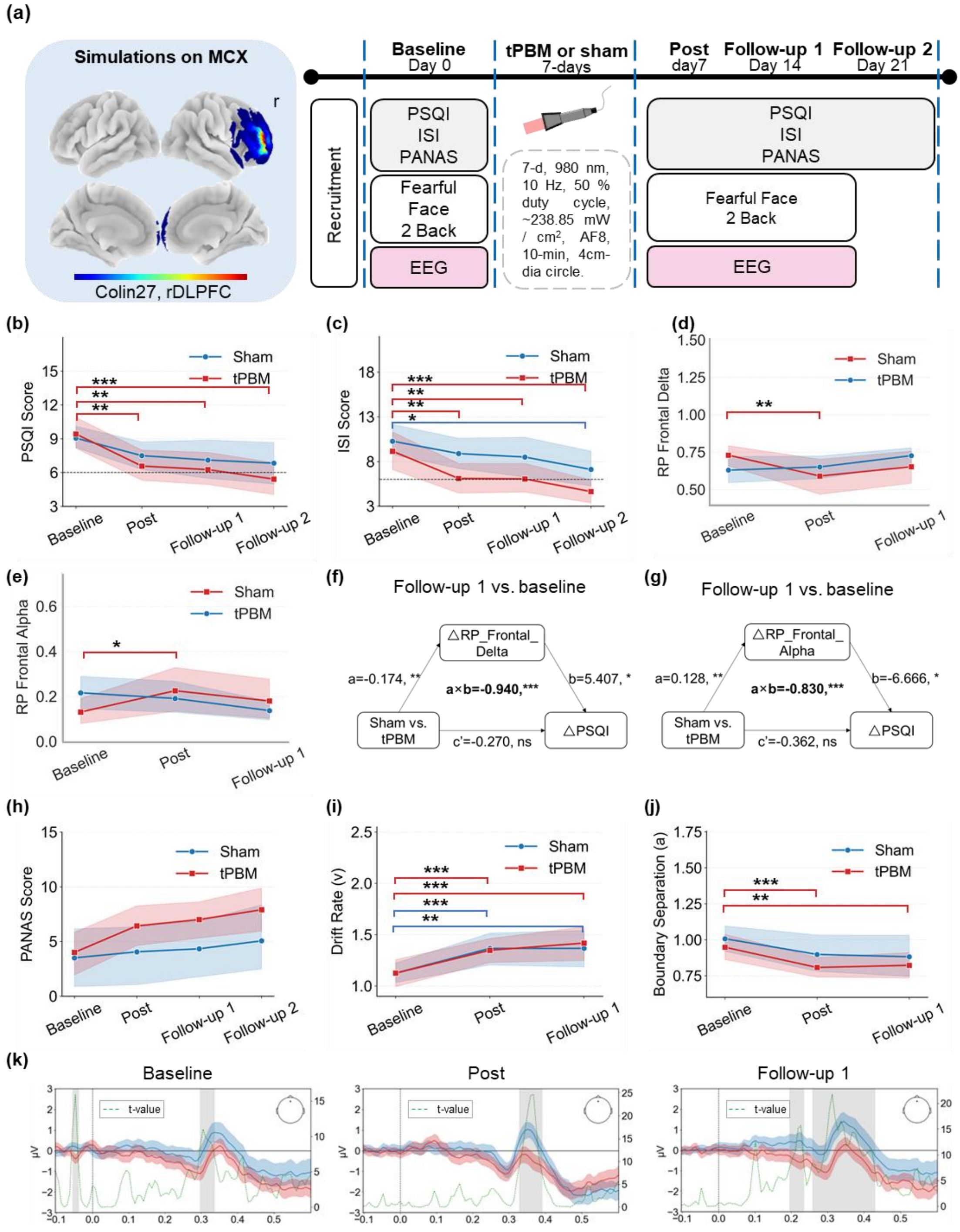

**Figure 1.** tPBM therapy for of insomnia disorder and potential neural mechanisms. (a) Photon flux simulation by Monte Carlo eXtreme (MCX) and experimental protocol. (b) Pittsburgh sleep quality index (PSQI) showed a significant main effect of Time ($\beta = -0.706$, SE = 0.191, $p = 2.274\times10^{-4}$) and a significant Group × Time interaction ($\beta = -0.526$, SE = 0.267, $p = 0.049$). (c) Insomnia severity index (ISI) showed a significant main effect of Time ($\beta = -0.989$, SE = 0.243, $p = 4.602\times10^{-5}$). (d) Relative delta power in frontal region (RP Frontal Delta) at Fz / F3 / F4 electrodes during closed eyes resting states, which showed a significant main effect of Time ($\beta = 0.054$, SE = 0.019, $p = 0.004$) and a significant Group × Time interaction ($\beta = -0.073$, SE = 0.027, $p = 0.006$). (e) Relative alpha power in frontal region (RP Frontal Alpha) at Fz / F3 / F4 electrodes during closed eyes resting states, which showed a significant main effect of Time ($\beta = -0.039$, SE = 0.015, $p = 0.008$) and a significant Group × Time interaction ($\beta = 0.048$, SE = 0.021, $p = 0.021$). (f)-(g) Mediation model with bootstrap demonstrating the effect of tPBM therapy on decrease the PSQI via increases in the relative delta power in frontal region, and decreases in the relative alpha power in frontal region. (h) There is no significant Group × Time interaction ($\beta = 0.732$, SE = 0.586, $p = 0.212$) was observed in positive and negative affect schedule (PANAS) score. (i) Drift diffusion model derived drift rate *v* is estimates for fearful face 2-back task, which showed a significant main effect of Time ($\beta = 0.121$, SE = 0.033, $p = 3.004\times10^{-4}$). (j) Drift diffusion model derived boundary separation *a* are estimates for fearful face 2-back task, which showed a significant main effect of Time ($\beta = -0.063$, SE = 0.022, $p = 0.004$). (k) Not-target event-related potential (ERP) waveforms at Fz electrodes, the gray shaded region highlights significant cluster corrected for multiple comparisons using the cluster-based permutation test (n = 1000 iterations). Note: Linear mixed model analysis with participant as a random intercept and fixed effects included Group (tPBM/sham), Time (baseline/post/follow-up 1/follow-up 2), and their interaction. Post hoc pairwise comparisons with Bonferroni correction were performed to explore the source of significant effects: within-group longitudinal changes (when Time or Interaction was significant) and between-group differences at each time point (when Group or Interaction was significant). *, $p < 0.05$, **, $p < 0.01$, ***, $p < 0.001$. Red and blue shaded areas represent the standard error of the mean.

This work was supported by the National Natural Science Foundation of China (Grant No. 62275210, 62506286), the National Leading Talent Program, the National Young Talent Program, the Postdoctoral Fellowship Program of CPSF (GZB20230561), the Key Research and Development Program of Shaanxi (Grant No. 2022GY-313), the Qinchuangyuan Scientists & Engineers of Shaanxi, the Xi'an Science and Technology Project (Grant No. 23ZDCYJSGG0026-2023), the Fundamental Research Funds for the Central Universities (Grant Nos. ZYTS23192, QTZX24079, and QTZX26160), the Xidian University Specially Funded Project for Interdisciplinary Exploration (Grant No. TZJH2024036), the Establishment Fund of the State Key Laboratory of Wakefullness/Sleep and Cognition(Chongqing Institute for Brain and Intelligence), the Science and Technology Development Fund (FDCT: 0062/2024/RIB2), and the Innovation Fund of Xidian University.

**Declaration of interest**

The authors declare that they have no known competing financial interests or personal relationships that could have appeared to influence the work reported in this paper.